\documentclass[seceq,preprint]{ptptex}

\usepackage{graphicx}
\usepackage{verbatim}
\usepackage{amsmath}
\usepackage{amssymb}
\usepackage{amsfonts}

\preprintnumber[2cm]{%<-- [..]: optional width of preprint # column.
YGHP-12-50}

\newcommand{\oper}[1]{\mathcal{#1}}
\newcommand{\comm}[2]{\left[ #1, #2 \right]}

\def\beq{\begin{eqnarray}}
\def\eeq{\end{eqnarray}}
\def\Tr{{\rm Tr}}
\def\Lag{\mathcal{L}}
\def\p{\partial}
\def\D{\mathcal{D}}

\newcommand {\non}{\nonumber\\}

\title{%        %You can use \\ for explicit line-break.Title%
Localization of matter fields and non-Abelian gauge fields
on domain walls
}

\author{%       %Use \scshape for the family name.
Masato \textsc{Arai}$^{a,b,}$\footnote{E-mail: masato.arai@gmail.com} 
Filip \textsc{Blaschke}$^{b,c,}$\footnote{E-mail: filip.blaschke@fpf.slu.cz}
Minoru \textsc{Eto}$^{d,}$\footnote{E-mail: meto@sci.kj.yamagata-u.ac.jp} 
and Norisuke \textsc{Sakai}$^{e,}$\footnote{E-mail: norisuke.sakai@gmail.com
}
}

\inst{%     %Affiliation, neglected when [addenda] or [errata].
$^a$Fukushima National College of Technology, Iwaki, Fukushima 970-8034, Japan\\
$^b$Institute of Experimental and Applied Physics, Czech Technical University in Prague, Horsk\'a 22, 128 00 Prague 2, Czech Republic\\
$^c$Institute of Physics, Silesian University in Opava, Bezru\v{c}ovo n\'am. 1150/13, 746~01 Opava, Czech Republic\\
$^d$Department of Physics, Yamagata University, Yamagata 990-8560, Japan\\
$^e$Department of Mathematics, Tokyo Woman's Christian University, \\
Tokyo 167-8585, Japan
}
\abst{
We propose a method for simultaneous localization of non-Abelian gauge fields and matter fields with minimal interaction on domain walls
using $U(N)_c$ gauge theory with  $SU(N)_{L}\times SU(N)_{R}\times U(1)_{A}$ 
flavor symmetry. Localization of non-Abelian fields is achieved using field-dependent gauge coupling. We find that effective Lagrangian up to
second order of derivatives for low
energy fluctuations resembles a chiral model from hadron physics with additional minimal interactions of pions with localized gauge fields together
with nonlinear interactions of moduli fields.  
This result provides a step towards a realistic model building 
of brane-world scenario using topological solitons. 
}

\begin{document}

\maketitle

\section{Introduction}\label{sc:introduction}
In the brane-world scenario  \cite{Horava:1995qa,ArDiDv,RaSu} it is assumed that all Standard Model (SM) fields are localized on 
(3+1)-dimensional defect called 3-brane, which is embedded in multidimensional space-time (bulk), where only graviton may propagate freely. 
To realize this scenario dynamically, we may use a topological soliton. A domain wall is the simplest soliton in (4+1)-dimensional theory.

Generally, bulk fields can have massless and massive modes.
It has been shown that massless modes localize naturally on the domain wall \cite{RuSh}.
Integrating out all massive modes, one obtains 
 effective field theory describing 
low energy interactions of these massless modes. 
However, in the context of field theory, localization of gauge fields, turned out to be difficult \cite{DuRu}.
It has been noted that the broken gauge symmetry 
in the bulk outside of the soliton inevitably makes 
the localized gauge field massive with the mass of the 
order of inverse width of the wall \cite{DaSh,Maru:2003mx}. 
Thus, to localize a massless gauge field, one needs to have 
the confining phase rather than the Higgs phase in the bulk 
outside of the soliton. 

Recently, a classical picture of the 
confinement \cite{Kogut:1974sn,Fukuda:1977wj} realized by  
position-dependent gauge coupling 
has been used to localize the non-Abelian 
gauge field on domain walls \cite{OhSa}. 
This position-dependent 
gauge coupling was naturally introduced on the domain wall 
background through a scalar-field-dependent gauge coupling 
function resulting from a cubic prepotential of supersymmetric 
gauge theories. This mechanism in its simplest form uses two copies of $U(1)_c$ gauge theory, where the difference of 
respective kink-like profiles of adjoint scalars in both sectors is taken as position-dependent gauge coupling. 

However, it is still an interesting problem to make localized matter fields and gauge fields interact.
The purpose of this work is to present a (4+1)-dimensional 
field theory model of localized massless matter fields 
minimally coupled to the non-Abelian gauge field which 
is also localized on the domain wall with the 
(3+1)-dimensional world volume. 
We derive the low-energy effective field theory of 
these localized matter and gauge fields and describe full non-linear interaction between moduli fields. 

To introduce non-Abelian flavor symmetry (to be gauged 
eventually) in the domain wall sector, we replace 
one of the two copies of the $U(1)_c$ gauge theory 
with the flavor symmetry $U(1)_L\times U(1)_R$ in 
\cite{EtFuNiOhSa}, by $U(N)_c$ gauge theory with 
the extended flavor (global) symmetry 
$SU(N)_L\times SU(N)_R\times U(1)_{A}$. 
By choosing the coincident domain wall solution for this 
domain wall sector, we obtain the 
maximal unbroken non-Abelian flavor symmetry group 
$SU(N)_{L+R}$ which is preserved in both left and 
right vacua outside of the domain wall. 
Therefore we can introduce gauge field for the 
(subgroup of) the flavor $SU(N)_{L+R}$ symmetry. 
In order to obtain the field-dependent gauge coupling 
function, for the gauge field localization mechanism 
 \cite{OhSa}, we also introduce a coupling between 
a scalar field and gauge field strengths 
inspired by supersymmetric gauge theories, 
although we do not make the model fully supersymmetric 
at present. 
This scalar-field-dependent gauge coupling function gives 
appropriate profile of position-dependent 
gauge coupling through the background domain wall solution. 
With this localization mechanism for gauge field, we find 
massless non-Abelian gauge fields localized on the domain 
wall. 

In what follows, we only concentrate on presenting our main result. Detailed discussion of involved calculations and 
other topic outside this main line is presented elsewhere \cite{arxiv}.   

\bigskip
\section{Gauge field localization}
\label{sec:Abelian-Higgs}

\subsection{The domain wall sector}

Let us briefly illustrate the localization mechanism for the gauge 
fields and the matter fields on the domain walls 
by using a simplest model in (4+1)-dimensional 
spacetime: two copies ($i=1, 2$) 
of $U(1)$ models, each of which has two flavors 
($L, R$) of charged Higgs scalar fields 
$H_i=(H_{iL}, H_{iR})$:
\beq
\Lag_i &=& - \frac{1}{4g_i^2}\left({\cal F}_{MN}^i\right)^2 
+ \frac{1}{2g_i^2}\left(\p_M\sigma_i\right)^2 
+ \left|\D_M H_i\right|^2 - V_i, \label{Abelian-Higgs} \\
V_i &=& \frac{g_i^2}{2}\left(|H_i|^2 - v_i^2\right)^2 
+ \left|\sigma_iH_i-H_iM_i\right|^2. 
\eeq
We use the metric $\eta_{MN}={\rm diag}(+,-,\cdots,-)$, 
$M,N=0,1,\cdots,4$. 
The Higgs field $H_i$ is charged with respect to the $U(1)_i$ 
gauge symmetry and the covariant derivative is given by
\beq
\D_M H_i = \p_M H_i + i w_M^i H_i,
\eeq
where $w_M^i$ is the $U(1)_i$ gauge field with the field strength 
\beq
{\cal F}_{MN}^i = \p_M w_N^i - \p_N w_M^i.
\eeq
Since we want domain walls, we will choose 
\beq
M_i = {\rm diag}\left(m_i,-m_i\right),\quad (m_i > 0),
\eeq
resulting in the $U(1)_{iA}$ flavor symmetry.
 We have included the neutral scalar fields $\sigma_i$ in this 
Abelian-Higgs model. 
Notice that the same coupling constant $g_i$ appears not only in front of the 
kinetic terms of the gauge fields and $\sigma_i$, 
but also as the the quartic coupling constant of $H_i$. 
We have taken this special relation among the coupling 
constants only to simplify computations. 
One may repeat the whole procedure in models with more 
generic coupling constants without changing essential results.

 There are two discrete vacua for each copy $i$ 
\beq
(H_{iL},H_{iR},\sigma_i) = \left(v_i,0,m_i\right),\ 
\left(0,v_i,-m_i\right).
\label{eq:vacua_AH}
\eeq

Thanks to the special choice of the coupling constants 
in $\Lag_i$ motivated by the supersymmetry, 
there are Bogomol'nyi-Prasad-Sommerfield (BPS) domain wall solutions in these models. 
Let $y$ be the coordinate of the direction orthogonal 
to the domain wall and we assume all fields depend 
on only $y$. 
%Then, by usual methods \cite{Isozumi:2004jc, Eto:2006pg}, the Hamiltonian can be written as follows 
%\beq
%{\cal H}_i &=& \frac{1}{2g_i^2}\left(\p_y \sigma_i 
%+ g_i^2\left(|H_i|^2 - v_i^2\right)\right)^2
%+ \left|\D_y H_i + \sigma_i H_i - H_i M_i\right|^2 \non
%&+& v_i^2 \p_y \sigma_i - \p_y
%\left((\sigma_i H_i - H_i M_i)H_i^\dagger\right) \non
%&\ge& v_i^2 \p_y \sigma_i - \p_y
% \left((\sigma_i H_i - H_i M_i)H_i^\dagger\right).
%\eeq
%Thus the Hamiltonian is bounded from below. 
%This bound is called Bogomol'nyi bound, and is saturated 
%when the following BPS equations are satisfied 
The BPS equations are
\beq
\p_y \sigma_i + g_i^2\left( |H_i|^2 - v_i^2\right) = 0,
\quad
\D_y H_i + \sigma_i H_i - H_i M_i=0.
\label{eq:BPSeq_AH}
\eeq
In order to obtain the domain wall solution 
interpolating the two vacua in Eq.~(\ref{eq:vacua_AH}), 
we impose the boundary conditions :
\begin{eqnarray}
(H_{iL},H_{iR},\sigma_i) &=& \left(0,v_i,-m_i\right), \; \; y=-\infty, 
\nonumber \\%\quad 
(H_{iL},H_{iR},\sigma_i) &=& \left(v_i,0,m_i\right), \; \quad y=\infty.
\label{eq:boundary_cond}
\end{eqnarray}
Tension $T_i$ of the domain wall is given by a topological 
charge as 
\beq
T_i = \int^\infty_{-\infty} dy\ \left[
v_i^2 \p_y \sigma_i - \p_y
\left((\sigma_i H_i - H_i M_i)H_i^\dagger\right)
\right]^\infty_{-\infty} = 2 m_i v_i^2.
\eeq

The second equation of the BPS equations (\ref{eq:BPSeq_AH}) 
can be solved by the moduli matrix formalism  
\cite{Isozumi:2004jc, Eto:2006pg} 
with the constant matrix (vector) $H_{i0} = (C_{iL},C_{iR})$ 
\beq
H_i = v_i e^{-\frac{\psi_i}{2}} H_{i0}e^{M_iy},\quad
\sigma_i + i w_i = \frac{1}{2} \p_y \psi_i.
\label{eq:mm_AH}
\eeq
For a given $H_{i0}$, the scalar function $\psi_i$ is determined 
by the master equation
\beq
\p_y^2\psi_i = 2 g_i^2v_i^2 \left(1 - e^{-\psi_i} 
H_{i0} e^{2M_iy} H_{i0}^\dagger\right).
\label{eq:master_Abelian-Higgs}
\eeq
%The asymptotic behavior of the field $\psi_i$ is determined 
%by the condition that the configuration reaches 
%the vacuum at left and right infinities:
%\beq
%\psi_i \to \log H_{i0} e^{2M_iy}H_{i0}^\dagger, 
%\quad |y|\to\infty.
%\eeq
There exists redundancy in the decomposition in 
Eq.~(\ref{eq:mm_AH}), which is called the $V$-transformation:
\beq
H_{i0} \to V_i H_{i0},\quad \psi_i \to \psi_i + 2 \log V_i,
\quad V_i \in \mathbb{C}^*.
\label{eq:V_Abelian-Higgs}
\eeq
%For example, a single domain wall solution centered at $y=0$ 
%can be generated by a moduli matrix
%\beq
%H_{i0} = (1,1).
%\eeq
%Then the master equation is 
%\beq
%\p_y^2 \psi_i = 2 g_i^2v_i^2 \left(1 - e^{-\psi_i}
%\left(e^{2m_iy} + e^{-2m_iy}\right)\right).
%\eeq
No analytic solutions for the master equation have been found 
for finite gauge couplings $g_i$. But we can take a strong gauge coupling limit $g_i \to \infty$ without losing any essential features
(see \cite{arxiv} for further details).
%so we must solve it numerically.
%The corresponding solution is shown in Fig.~\ref{fig:dw_sample}.
%\begin{figure}[ht]
%\begin{center}
%\begin{tabular}{cc}
%\includegraphics[height=4cm]{dw_sample.eps} 
%\includegraphics[height=4cm]{dw_sample2.eps}  
%\end{tabular}
%\caption{The left panel shows profiles of $H_{iL}$ (solid line), 
%$H_{iR}$ (long-dashed line), 
%and $\sigma$ (dashed-line) with finite 
%gauge coupling ($g_i=0.5$). 
%The right panel shows a plot of $\sigma$: 
%dashed curve for finite ($g_i=0.5$) gauge coupling 
%and solid curve for strong gauge coupling ($g_i=\infty$). 
%The other parameters are $m_i=v_i=1$. }
%\label{fig:dw_sample}
%\end{center}
%\end{figure}
The generic solutions of the domain wall are generated by 
the generic moduli matrices (after fixing the $V$-transformation)
\beq
H_{i0} = \left(C_{iL}, C_{iR}\right),\quad C_{iL}, C_{iR} \in \mathbb{C}^*.
\eeq
The complex constants $C_{iL}, C_{iR}$ are free parameters 
containing the moduli parameters of the BPS solutions.
The moduli parameter can be defined by 
\beq
C_i \equiv \sqrt{\frac{C_{iR}}{C_{iL}}} = e^{i\alpha_i} e^{m_i y_i}.
\label{eq:sol_sample} 
\eeq
The other degrees of freedom in $C_{iL}, C_{iR}$ can be 
eliminated by the $V$-transformation in Eq.~(\ref{eq:V_Abelian-Higgs}) 
and has no physical meaning. 
Then, in the strong gauge coupling limit, the master equation is found to be (where we used coincident walls set-up $H_{i0} = (1,1)$)
\beq
1 = e^{-\psi_i}
\left(e^{2m_i(y-y_i)} + e^{-2m_i(y-y_i)}\right)\,,
\label{eq:master_Abelian-Higgs2}
\eeq
with the solution
\beq
\sigma_i = m_i \frac{1-|\phi_i|^2}{1+|\phi_i|^2} 
= m_i \tanh 2m_i(y-y_i)\,.
\eeq
It is obvious that the real parameter $y_i$ is the 
translational moduli of the domain wall.
The other parameter $\alpha_i$ is an internal moduli 
which is the Nambu-Goldstone (NG) mode associated with
the $U(1)_{iA}$ flavor symmetry spontaneously broken 
by the domain walls.

Let us next derive the low energy effective theory on 
the domain wall.
We integrate all the massive modes while keeping the 
massless modes. 
We use the so-called moduli approximation where the 
dependence on (3+1)-dimensional spacetime coordinates 
comes into the 
effective Lagrangian only through  the moduli fields:
\beq
C_i \to C_i(x^\mu),\quad
\phi_i(y) \to \phi_i(y,C_i(x^\mu)) = C_i(x^\mu)^2e^{-2m_iy}.
\eeq
The effective Lagrangian for the moduli field $C_i(x^\mu) = e^{i\alpha} e^{m_iy_i}$ is given by
%can be obtained by plugging this into the Lagrangian $\Lag_i$ 
%and integrate it over $y$. 
%This can be done explicitly as follows.
%\beq
%\Lag_{i,{\rm eff}} =  \int^\infty_{-\infty} dy 
%\frac{v_i^2}{\left(|C_i|^{-2}e^{2m_iy}+|C_i|^2 e^{-2m_iy}\right)^2} 
%\frac{\left|\p_\mu C_i^2\right|^2}{|C_i^2|^2}
%= \frac{v_i^2}{4m_i } 
%\frac{\left|\p_\mu C_i^2\right|^2}{\left|C_i^2\right|^2}.
%\label{eq:eff_Abelian-Higgs0}
%\eeq
%If we take $C_i = e^{i\alpha} e^{m_iy_i}$, the effective Lagrangian is given by
\beq
\Lag_{i,{\rm eff}} 
= \frac{2m_iv_i^2}{2}(\p_\mu y_i)^2 
+ \frac{v_i^2}{m_i}(\p_\mu \alpha_i)^2,
\label{eq:effLagAbelian}
\eeq
where energy of soliton solution is neglected since it does not contribute
to dynamics of moduli.
Note that $2m_iv_i^2$ is precisely the domain wall tension.
This is the free field Lagrangian.

%Although we have derived this effective Lagrangian in the 
%strong gauge coupling limit, we can obtain the same 
%Lagrangian in the finite gauge coupling constant. 
%In other words, the effective  Lagrangian cannot 
%distinguish the infinite versus finite coupling cases 
%at least in the quadratic order of the derivative expansion.

%\smallskip
%\subsection{Localization of the Abelian gauge fields 
%\label{sec:localization_Abelian}}

%In the previous subsection, we have seen 
%that the NG modes of the translation and $U(1)$ global 
%symmetry are the only massless modes in the Abelian-Higgs 
%model. 
%They are localized on the domain wall. 
%There are no massless gauge field on the domain wall and 
%all the modes contained in the gauge field are massive. 
%The mass of the lightest mode of the gauge field is of the 
%order of the inverse of the width of the domain wall, 
%since the bulk outside of the domain wall is in the Higgs 
%phase. 
%The low energy effective Lagrangian for the massless fields 
%is obtained after integrating out the massive modes including 
%gauge fields. 

In order to obtain the massless gauge field to be localized 
on the domain wall, we need a new gauge symmetry which is 
unbroken in the bulk.
Recently, a new mechanism was proposed to localize 
gauge fields on domain walls \cite{OhSa}.

A key ingredient is the so-called dielectric coupling 
constant \cite{Kogut:1974sn,Fukuda:1977wj}. 
%To illustrate the new localization mechanism, 
Let us 
introduce a new $U(1)$ gauge field $a_M$ which we wish to 
localize on the domain wall. 
Since this gauge symmetry should be unbroken in the bulk, 
we consider the case where all the Higgs fields are neutral 
under this newly introduced $U(1)$ gauge symmetry. 
The gauge field $a_M$ is assumed to couple to the neutral 
scalar fields $\sigma_i$ only in the following particular 
combination 
\beq
\Lag = \Lag_1 + \Lag_2 - \frac{\lambda}{2}
\left(\frac{\sigma_1}{m_1} 
- \frac{\sigma_2}{m_2}\right)\left({\cal G}_{MN}\right)^2,
\label{eq:lag_Abelian-Higgs_local}
\eeq
where a real constant $\lambda$ with the unit mass 
dimension, in accordance with the (4+1)-dimensional 
spacetime and the field strength is defined by 
\begin{equation}
{\cal G}_{MN} = \p_M a_N - \p_N a_M. 
\end{equation}
The field-dependent gauge coupling function is given by
\beq
\frac{1}{4e^2(\sigma)} = \frac{\lambda}{2}
\left(\frac{\sigma_1}{m_1} - \frac{\sigma_2}{m_2}\right), 
\label{eq:coupling_Abelian-Higgs}
\eeq
which depends on the position $y$ through fields $\sigma_i$.  
Thus the field-dependent gauge coupling function 
$e(\sigma)$ plays the role 
of the dielectric coupling constant.
The form of Eq.~(\ref{eq:coupling_Abelian-Higgs}) is chosen so that the gauge interaction becomes strongly coupled in the bulk
($\sigma_i \to \pm m_i$ 
as $y \to \pm \infty$).

%Let us again consider a double copy of domain walls as a 
%background configuration 
%in the Abelian-Higgs model in Eq.~(\ref{eq:lag_Abelian-Higgs_local}). 
Since our Lagrangian has no term 
linear in $a_M$, the equations of motion for 
$a_M$ is trivially solved by $a_M=0$, and the rest of 
the equations of motion are explicitly the same as 
those in the previous subsection. 
Therefore the domain wall solution in the previous 
subsection together with $a_M=0$ is still a solution 
of the equations of motion. 
%\beq
%\Lag_{\rm eff} = \Lag_{1,{\rm eff}} + \Lag_{2,{\rm eff}} 
%- \frac{1}{4e_4^2} ({\cal G}_{\mu\nu})^2,
%\eeq
%except for the additional kinetic term 
%(the last term) of the (3+1)-dimensional gauge 
%field $w_{\mu}$, which is the zero mode 
%($y$-independent mode) of 
%the (4+1)-dimensional field $w_{\mu}$. 
Thus, the (3+1)-dimensional gauge coupling constant is given by 
\beq
 \frac{1}{4e_4^2} = \frac{\lambda}{2}\int^\infty_{-\infty} 
dy\ \Bigl(\frac{\sigma_1}{m_1} - \frac{\sigma_2}{m_2}\Bigr)
 = \frac{\lambda}{4}\left[\frac{\psi_1}{m_1} 
- \frac{\psi_2}{m_2}\right]^\infty_{-\infty} 
= \lambda(y_2-y_1),
\label{eq:integral4Dcoupling}
 \eeq
 where we %have chosen $m_1=m_2$ for simplisity and 
have used the asymptotic behavior 
$\psi_i \to \log 2\cosh 2m_i(y-y_i)$ as 
$|y|\to \infty$.
% Note that this result is again independent of the gauge 
%couplings $g_i$ in the domain wall sector.
% In summary, the low energy effective Lagrangian is
The low energy effective Lagrangian is found to be
 \beq
 \Lag_{\rm eff} = 
 \sum_{i=1,2}\left[\frac{2m_iv^2}{2}(\p_\mu y_i)^2 
+ \frac{v_i^2}{m_i}
(\p_\mu \alpha_i)^2\right] 
 - \lambda (y_2-y_1) ({\cal G}_{\mu\nu})^2.
 \label{eq:eff_Abelian-Higgs}
 \eeq
%Now we separate the quantum fields (fluctuations) from the 
%classical background moduli parameters by
%\beq
%y_i(x^\mu) = y_i^0 + \delta y_i,\quad
%\alpha_i(x^\mu) = \alpha_i^0 + \delta \alpha_i.
%\label{eq:separate_Abelian-Higgs}
%\eeq
%Then the effective Lagrangian 
%up to the second order of the small quantum fluctuations 
%is given by
% \beq
% \Lag_{\rm eff}(y_i^0,\alpha_i^0) = 
% \sum_{i=1,2}\left[\frac{2m_iv_i^2}{2}(\p_\mu \delta y_i)^2 
%+ \frac{v_i^2}{m_i}(\p_\mu \delta \alpha_i)^2\right] 
% - \lambda(y_2^0-y_1^0) ({\cal G}_{\mu\nu})^2,
% \label{eq:eff_Abelian-Higgs}
% \eeq
%We note that the massless gauge field $a_\mu$ has a 
%positive finite gauge coupling squared %\footnote{ 
%Here we are content with the fact that the positivity of 
%the gauge kinetic term is assured at least in finite region 
%of moduli space, instead of just at a point. 
%However, it is possible to make a more economical model 
%where one has less moduli, and the positivity of the 
%gauge kinetic term is assured\cite{OhSa}. 
%} 
%$1/(4\lambda(y_2^0-y_1^0))$ provided $y_2^0-y_1^0>0$. 

Although we succeeded in localizing the massless $U(1)$ 
gauge field $a_\mu$ on the domain walls, the Lagrangian 
Eq.~(\ref{eq:eff_Abelian-Higgs}) has no charged matter fields 
minimally coupled with the localized gauge field $a_\mu$. 
In the next section, we will give a model with 
a non-Abelian global symmetry whose unbroken subgroup 
can be gauged to yield massless localized gauge fields 
on the domain wall. 
%%\end{comment}

\bigskip
\section{The chiral model 
\label{sec:chiral model}}

In this section we study domain walls in the chiral model 
which is a natural extension of the Abelian-Higgs model 
in the previous section. 
%This chiral model leads to two important consequences 
%1) massless non-Abelian gauge fields are localized on the 
%domain wall and moreover 
%2) the scalar fields which are non-trivially interacting 
%are also localized on the domain walls. 

%\smallskip
%\subsection{The domain walls in the chiral model}

%As a natural extension of the domain wall sector in the 
%previous section, we 
Let us consider the Yang-Mills-Higgs model 
with $SU(N)_{c}\times U(1)$ gauge symmetry with 
$S[U(N)_{L}\times U(N)_{R}] = SU(N)_{L}\times SU(N)_{R} 
\times U(1)_{A}$ flavor symmetry \cite{ShYu2,EtFuNiOhSa}. 
To localize the gauge field in a simple manner, 
we again introduce two sectors $\Lag_1$ and $\Lag_2$, but 
only the former is extended to 
Yang-Mills-Higgs system and the latter is the same form 
as in (\ref{Abelian-Higgs}) with $i=2$.
%The second sector couples to the first sector through 
%the coupling as described in 
%(\ref{eq:lag_Abelian-Higgs_local}) and after gauging the flavor symmetry 
% it plays a role as localization of gauge fields, combined 
%with the first sector.
The matter contents are summarized in Table \ref{table:SY}. 
Since the presence of two factors of $SU(N)$ global 
symmetry resembles the chiral symmetry of QCD, we call 
this Yang-Mils-Higgs system as the chiral model. 
\begin{table}
\begin{center}
\begin{tabular}{c|cccccccc}
\hline
  & $SU(N)_{c}$ & $U(1)_1$ & $U(1)_2$ & $SU(N)_{L}$ 
& $SU(N)_{R}$ & $U(1)_{1A}$ & $U(1)_{2A} $ & mass\\ \hline
$H_{1L}$ & $\square$ & 1 & 0 & $\square$ & {\bf 1} 
& 1 & 0 & $m_1{\bf 1}_{N}$\\ 
$H_{1R}$ & $\square$ &  1 & 0 & {\bf 1} 
& $\square$ & $-1$ & 0 & $- m_1{\bf 1}_{N}$ \\ 
$\Sigma_1$ & ${\rm adj}\oplus{\bf 1}$ & 0 & 0 & {\bf 1} 
& {\bf 1} & 0 & 0 & 0\\
$H_{2L}$ & {\bf 1} & 0 & 1 & {\bf 1} 
& {\bf 1} & 0 & $1$  & $m_2$\\ 
$H_{2R}$ & {\bf 1} & 0 & 1 & {\bf 1} 
& {\bf 1}& 0 & $-1$ & $- m_2$ \\ 
$\Sigma_2$ & {\bf 1} & 0 & 0 & {\bf 1} 
& {\bf 1} & 0 & 0 & 0\\
\hline
\end{tabular}
\end{center}
\caption{Quantum numbers of the domain wall sectors in the 
chiral model. 
%Yang-Mills-Higgs model.
}
\label{table:SY}
\end{table}

The Lagrangian is given by
\beq
\Lag &=& \Lag_1+\Lag_2, 
\label{eq:Lag_chi_sum}\\
\Lag_1 &=& \Tr\left[-\frac{1}{2g_1^2}(F_{1MN})^2 
+ \frac{1}{g_1^2}(\D_M\Sigma_1)^2 
+ \left|\D_M H_1\right|^2 \right] - V_1,
\label{eq:Lag_chi}\\
V_1 &=& \Tr\left[ \frac{g_1^2}{4} \left(H_1 H_1^\dagger 
- v_1^2 {\bf 1}_{N}\right)^2 + \left|\Sigma_1 H_1
-H_1M_1\right|^2\right],
\label{eq:YM_Higgs_Lag}
\eeq
with $H_1 = \left(H_{1L},\ H_{2L}\right)$.
$\Lag_2$ is the same form as (\ref{Abelian-Higgs}) with $i=2$.
Gauge fields of $U(N)_c=(SU(N)_c\times U(1)_1)/Z_{\bf N}$ 
are denoted as $W_{1M}$, and adjoint scalar as 
$\Sigma_{1}$. 
The covariant derivative and the field strength are denoted as 
$\D_M\Sigma_1 = \p_M \Sigma_1 + i \left[W_{1M}, \Sigma_1\right]$, 
$\D_M H_1 = \p_M H_1 + i W_{1M} H_1$, 
and $F_{1MN} = \p_M W_{1N} - \p_N W_{1M} 
+ i \left[W_{1M},W_{1N}\right]$. 
The mass matrix is given by 
$M_1 = {\rm diag}\left(m_1 {\bf 1}_N, -m_1{\bf 1}_N\right)$. 
Let us note that the chiral model reduces to the Abelian-Higgs 
model in the limit of $N\to1$, by deleting all the 
$SU(N)$ groups. 

The role of the second sector is to provide the localization mechanism in the similar way as in the previous section.
Therefore, the remaining discussion will be focused solely on the 
first sector ($i=1$) and we suppress the index $i=1$.
%The symmetry transformations act on the fields as
%\beq
%H = \left(H_{L},H_{R}\right) &\to& U_{c}\left(H_{L},H_{R}\right)
%\left(
%\begin{array}{cc}
%U_{L}e^{i\alpha} & \\
%& U_{R}e^{-i\alpha}
%\end{array}
%\right),
%\label{eq:transf_H}
%\\
%\Sigma &\to& U_{c}\Sigma U_{c}^\dagger,
%\label{eq:transf_sigma}
%\eeq
%with $U_{c} \in U(N)_{c}$, $U_{L} \in SU(N)_{L}$, 
%$U(N)_{R} \in SU(N)_{R}$ and $e^{i\alpha} \in U(1)_{A}$.

There exist $N+1$ vacua in which the fields develop the 
following VEV
\beq
H &=& (H_{L}, H_{R}) = v\left(
\begin{array}{cc|cc}
{\bf 1}_{N-r} & & {\bf 0}_{N-r} & \\
& {\bf 0}_{r} & & {\bf 1}_{r}
\end{array}
\right),\\
\Sigma&=& m \left(
\begin{array}{cc}
{\bf 1}_{N-r} & \\
& -{\bf 1}_{r}
\end{array}
\right),
\eeq
with $r=0,1,2,\cdots, N$.
We refer these vacua with the label $r$. 
In the $r$-th vacuum, both the local gauge symmetry $U(N)_c$ 
and the global symmetry are broken, but a diagonal global 
symmetries are unbroken (color-flavor-locking) 
\begin{eqnarray}
&&U(N)_c \times SU(N)_L \times SU(N)_R \times U(1)_{A} 
\to 
\nonumber \\
&&SU(N-r)_{L+c}\times SU(r)_L \times 
SU(r)_{R+c}\times SU(N-r)_R\times U(1)_{A+c}.
\label{eq:symmetry_r_vacuum}
\end{eqnarray}

%As in the Abelian-Higgs model, 
The BPS equations for the 
domain walls can be obtained as:
%through the 
%Bogomol'nyi completion of the energy density with the 
%assumption that all the fields depend on only the
%fifth coordinate $y$ and $W_{\mu} = 0$:
%\beq
%{\cal H} &=& \Tr\left[
%\frac{1}{g^2}\left(\D_y\Sigma 
%- \frac{g^2}{2}\left(v^2 {\bf 1}_N 
%- HH^\dagger\right)\right)^2
%+\left|\D_yH+\Sigma H-HM\right|^2
%\right]\non
%&+& \p_y \left\{ \Tr \left[v^2 \Sigma 
%- \left(\Sigma H-HM\right)H^\dagger\right]\right\} \non
%&\ge&  \p_y \left\{ \Tr \left[v^2 \Sigma 
%- \left(\Sigma H-H M \right)H^\dagger\right]\right\}.
%\eeq
%This bound is saturated when the following BPS equations 
%are satisfied 
\beq
\D_y\Sigma - \frac{g^2}{2}\left(v^2 {\bf 1}_{N} 
- H H^\dagger\right) = 0,\label{eq:BPS_SY1}\\
\D_y H+\Sigma H-H M = 0.\label{eq:BPS_SY2}
\eeq
The tension of the domain wall is given by 
\beq
T &=& \int^\infty_{-\infty} dy \ \p_y 
\left\{\Tr \left[v^2\Sigma 
- \left(\Sigma H-H M\right)H^\dagger\right]\right\} \non
&=& v^2 \; {\rm Tr} \left[\Sigma(+\infty) 
- \Sigma(-\infty)\right].
\label{eq:tension_SY}
\eeq 

Let us concentrate on the domain wall which connects the 
0-th vacuum at $y\to \infty$ and the $N$-th vacuum at 
$y \to -\infty$. Its tension can be read as 
\beq
T = 2N v^2m,
\eeq
from
Eq.~(\ref{eq:tension_SY}). 
Since there are $N+1$ possible vacua, the maximal number of 
walls is $N$ at various positions. 
The simplest domain wall solution corresponding to the 
coincident walls is given by making an 
ansatz that $H_{L}$, $H_{R}$, $\Sigma$ and $W_{y}$ 
are all proportional to the unit matrix. 
Then the BPS equations (\ref{eq:BPS_SY1}) and 
(\ref{eq:BPS_SY2}) can be identified with 
the BPS equations in Eq.~(\ref{eq:BPSeq_AH}) in  the 
Abelian-Higgs model.
Thus the domain wall solution can be solved as
\begin{align}
H_{L} &= v e^{-\frac{\psi}{2}}e^{m y}~{\bf 1}_{N},\label{eq:sol1_SY}\\
H_{R} &= v e^{-\frac{\psi}{2}}e^{-m y}~{\bf 1}_{N},\label{eq:sol2_SY}\\
\Sigma + i W_{y} &= \frac{1}{2}\p_y\psi{\bf 1}_{N}\label{eq:sol3_SY},
\end{align}
where $\psi$ is the solution of the master equation 
(\ref{eq:master_Abelian-Higgs}) in the Abelian-Higgs model.
Eq.(\ref{eq:symmetry_r_vacuum}) shows that the unbroken 
global symmetry for $N$-th vacuum ($H_{L} = 0$, 
$H_{R} = v{\bf 1}_{N}$ and $\Sigma = -m{\bf 1}_{N}$) 
at the left infinity $y \to -\infty$ is 
$SU(N)_L \times SU(N)_{R+c} \times U(1)_{A+c}$, 
whereas that for the $0$-th vacuum ($H_{L} = v{\bf 1}_{N}$, 
$H_{R} = 0$ and $\Sigma = m{\bf 1}_{N}$) 
at the right infinity $y\to\infty$ is 
$SU(N)_{L+c} \times SU(N)_R \times U(1)_{A+c}$.

The domain wall solution further breaks these unbroken 
symmetries because it interpolates the two vacua.
%The breaking pattern by the domain wall is \footnote{
%The unbroken generators of $U(1)_{A+c}$ for $r$-th 
%vacuum contains different combination of 
%$U(N)_c$ generators depending on $r$. 
%Therefore the right and left vacua preserve actually 
%different $U(1)_{A+c}$, and the wall solution does not 
%preserve any of these $U(1)_{A+c}$. 
%}
\beq
U(N)_c \times SU(N)_L \times SU(N)_R \times U(1)_{A} 
\to SU(N)_{L + R+c}.
\eeq
%One can easily verify that $SU(N)_{L+R+c}$ is unbroken by 
%the domain wall solution in Eqs.~(\ref{eq:sol1_SY}) -- 
%(\ref{eq:sol3_SY}).
This spontaneous breaking of the global symmetry gives 
NG modes on the domain wall as massless degrees of freedom 
valued on the coset similarly to the chiral symmetry breaking 
in QCD : 
\beq
\frac{SU(N)_{L} \times SU(N)_R}{SU(N)_{L + R+c}} 
\times U(1)_{A}.
\label{eq:NGmodes_wall}
\eeq
Since our model can be embedded into a supersymmetric 
field theory \cite{arxiv}, these NG modes ($U(N)$ chiral fields) appear as complex scalar fields
accompanied with additional $N^2$ pseudo-NG modes\footnote{
One of them is actually a genuine NG mode corresponding to 
the broken translation. }.

%\subsection{Localization of the matter fields}
%\label{sec:chiral}

%In the remainder of this subsection, 
We will now give 
the low-energy effective Lagrangian 
on the domain walls where the massless moduli fields (the matter fields) are localized.
The best way to parametrize these massless moduli 
fields is to use the moduli matrix 
formalism \cite{Isozumi:2004jc, Eto:2006pg} 
\begin{align}
H_{L} & = v e^{my}S^{-1}\,,\label{eq:solU1_SY}\\
H_{R} & = v e^{-my} S^{-1}e^{\phi}\,,\label{eq:solU2_SY}\\
\Sigma + i W_{y} & = S^{-1}\partial_y S\,,
\label{eq:solU3_SY}
\end{align}
where $S\in GL(N,{\bf C})$ and $\Omega=S S^\dagger$ is the 
solution of the following 
master equation 
\begin{eqnarray}
\partial_y\left( \Omega^{-1}\partial_y \Omega\right) = g^2v^2 
  \left(\mathbf{1}_N - \Omega^{-1}\Omega_0\right)\,, 
\label{eq:wll:master-eq-wall}
\end{eqnarray}
where
\begin{equation*}
\Omega_0 
= e^{2my}\mathbf{1}_N+e^{-2my}e^{\phi}e^{\phi^{\dagger}}\,.
\label{eq:omega0}
\end{equation*}
We have used the $V$-transformation to identify 
the moduli $e^\phi$, which is a complex $N$ by $N$ matrix. 
It can be parametrized by an $N \times N$ hermitian matrix 
$\hat x$ 
and a unitary matrix $U$ as \cite{EtFuNiOhSa} 
\begin{eqnarray}
e^{\phi} = e^{\hat x}U^{\dagger}, 
\label{eq:xu_decomp}
\end{eqnarray}
where $U$ is nothing but the $U(N)$ chiral fields associated with
the spontaneous symmetry breaking Eq.~(\ref{eq:NGmodes_wall}) and
$\hat x$ is the pseudo-NG modes whose existence we promised above.

In the strong gauge coupling limit $g \to \infty$, 
a solution of master equation is simply $\Omega = \Omega_0$. 
After fixing the $U(N)_c$ gauge, 
we obtain
\begin{eqnarray}
S=e^{\hat x/2}\sqrt{2\cosh (2my-\hat x)}\,. 
\end{eqnarray} 
Let us denote, for brevity 
\begin{equation}
\hat y= 2my-\hat x\,, 
\end{equation}
the Higgs fields are then given as 
\begin{align}
H_{L} & = v \frac{e^{\hat y/2}}{\sqrt{2\cosh \hat y}}\,,
\label{eq:solU1_SY3}\\
H_{R} & = v \frac{e^{-\hat y/2}}{\sqrt{2\cosh \hat y}} 
U^\dagger\,.
\label{eq:solU2_SY3}
\end{align}
From this solution, one can easily recognize that eigenvalues of $\hat x$
correspond to the positions of the $N$ domain walls in the $y$ direction.
Now we  promote moduli parameters $\hat x$ and 
$U$ to fields on the domain wall world volume, namely 
functions of world volume coordinates $x^\mu$. 
We plug the domain wall solutions 
$H_{L,R}(y; \hat x(x^\mu),U(x^\mu))$ 
into the original Lagrangian $\Lag$ in 
Eq. (\ref{eq:Lag_chi}) at $g \to \infty$ 
and pick up the terms quadratic in the derivatives. 
Thus the low energy effective Lagrangian is given by
\beq
\Lag_{{\rm eff}} = \int_{-\infty}^{\infty} dy\,\Tr\left[
\partial_\mu H_{L}\partial^\mu H_{L}^\dagger 
+ \partial_\mu H_{R}\partial^\mu H_{R}^\dagger
-v^2 W_\mu W^\mu
\right]\,,
\eeq
where
\beq
 W_\mu={i \over 2v^2}\left[\partial_\mu H_LH_L^\dagger 
-H_L\partial_\mu H_L^\dagger+(L\leftrightarrow R)\right].
\eeq
Here we have eliminated the massive gauge field $W_\mu$ 
by using the equation of motion.
Let us next introduce the gauge fields which are to be 
massless and localized on the domain walls. 
Since associated gauge 
symmetry should not be broken by the domain walls, we can gauge only unbroken 
symmetry $SU(N)_{L+R+c}$ itself or its subgroup. 
Let us gauge $SU(N)_{L+R}\equiv SU(N)_V$ and let $A_{\mu}^a$ be the $SU(N)_{L+R}$ gauge field.
The  Higgs fields are in the bi-fundamental representation of $U(N)_c$ and $SU(N)_{L+R}$.
The covariant derivatives of the Higgs fields are modified by
\beq
\tilde\D_M H_{1L} = \p_M H_{1L} + i W_{1M} H_{1L} 
- i H_{1L}A_{M},\label{eq:covd1}\\
\tilde\D_M H_{1R} = \p_M H_{1R} + i W_{1M} H_{1R} 
- i H_{1R}A_{M}\label{eq:covd2}.
\eeq
%The quantum numbers are summarized in Table \ref{table:gaugedSY}.
%\begin{table}
%\begin{center}
%\begin{tabular}{c|ccccccc}
%\hline
%  & $SU(N)_{c}$& $U(1)_1$ & $U(1)_2$ &$SU(N)_{V}$  & $U(1)_{1A}$ & $U(1)_{2A}$ & mass\\ \hline
%$H_{1L}$ & $\square$ & 1 & 0 & $\square$ & 1  & 0 & $m_1{\bf 1}_{N}$\\ 
%$H_{1R}$ & $\square$ & 1 & 0 & $\square$ & $-1$ & 0 & $- m_1{\bf 1}_{N}$ \\ 
%$\Sigma_1 $ & ${\rm adj}\oplus{\bf 1}$  & 0 & 0 & {\bf 1} & 0 & 0 & 0 \\
%$H_{2L}$ & {\bf 1} & 0 & 1 & {\bf 1} & 0 & 1  & $m_2$\\ 
%$H_{2R}$ & {\bf 1} & 0 & 1 & {\bf 1} & 0 & $-1$ & $- m_2$ \\ 
%$\Sigma_2 $ & {\bf 1} & 0 & 0 & {\bf 1} & 0 & 0 & 0  \\

%\hline
%\end{tabular}
%\end{center}
%\caption{Quantum numbers of the domain wall sectors in gauged chiral model}
%\label{table:gaugedSY}
%\end{table}

We now introduce a field-dependent gauge coupling function 
$g^2(\Sigma)$ for $A_{M}$, which is inspired by 
the supersymmetric model in Ref.\cite{OhSa}. 
\beq
\frac{1}{2e^2(\Sigma)} = \frac{\lambda}{2}
\left(\frac{\Tr\Sigma_1}{Nm_1} 
- \frac{\Sigma_2}{m_2}\right).
\label{Lag_gChi2}
\eeq
The Lagrangian is given by 
\beq
\Lag = \tilde{\Lag_1}+\Lag_2
 - \frac{1}{2e^2(\Sigma)}
\Tr\left[G_{MN}G^{MN}\right]. 
\label{Lag_gChi}
\eeq
The $\tilde \Lag_1$ in Eq.~(\ref{Lag_gChi}) 
is given by Eq.~(\ref{eq:Lag_chi}) where the covariant 
derivatives are replaced with those in Eqs.~(\ref{eq:covd1}) 
and (\ref{eq:covd2}).

We first wish to find the domain wall solutions in this 
extended model.
As before, we make ansatz that all the fields depend on 
only $y$ and $W_{\mu} = A_{\mu} =0$.
Let us first look on the E.O.M. of the new gauge field $A_M$.
It is of the form
\beq
\D_MG^{MN} = J^N,
\label{eq:eom_A}
\eeq
where $J_{M}$ stands for the current of $A_{M}$.
Note that the current $J_{M}$ is zero, by definition, if 
we plug the domain wall solutions
in the chiral model before gauging the $SU(N)_{L+R}$. 
This is because the domain wall 
configurations do not break $SU(N)_{L+R}$.
Therefore, $A_{M}=0$ is a solution of Eq.~(\ref{eq:eom_A}).

Then, we are left with equation of motion with $A_{M}=0$ 
which are identical to those in the ungauged chiral model 
in the previous subsections. 
Therefore the gauged chiral model admits 
the same domain wall solutions 
as those (Eqs.~(\ref{eq:solU1_SY3}) and (\ref{eq:solU2_SY3})) in the ungauged chiral model.

The next step is to derive the low energy effective theory on 
the domain wall world-volume 
in the moduli approximation as before.
%Again, we promote the moduli parameters as the fields on the 
%domain wall world-volume 
%and pick up the terms up to the quadratic order of the 
%derivative $\p_\mu$.
%Similarly to Sec.~\ref{sec:chiral}, we utilize 
%the strong gauge coupling limit $g_i\to \infty$, 
%to simplify the computation without changing the final 
%result. 
%Let us emphasize that we keep the field-dependent 
%gauge coupling function $e(\Sigma)$ finite. 
%The spectrum of massless NG modes is unchanged 
%by switching on the $SU(N)_{L+R}$ gauge interactions 
%\footnote{
%Tree level mass spectra are unchanged even though the chiral 
%symmetry $SU(N)_L \times SU(N)_R$ is 
%broken by the $SU(N)_{L+R}$ gauge interactions. 
%}. 
%
We just repeat the similar computation presented above.
%  to those in Sec.~\ref{sec:chiral}. 
%Again we shall focus on the first sector ${\cal L}_1$ and suppress the index $i=1$ of fields.
%Since color gauge fields $W_\mu$ becomes auxiliary fields 
%and eliminated through their equations of motion, 
It is 
convenient to define the covariant derivative only for the 
flavor ($SU(N)_{L+R}$) gauge interactions as 
\beq
\hat D_\mu H = \p_\mu H - i H A_{\mu}.
\eeq
Then we obtain the effective Lagrangian of the first sector as
\begin{eqnarray}
\oper{L}_{1,\rm eff} &=& \int_{-\infty}^{\infty} dy \,
\mathrm{Tr}\Bigl[\hat D_{\mu}H_{L}(\hat D^{\mu}H_{L})^{\dagger}
+\hat D_{\mu}H_{R}(\hat D^{\mu}H_{R})^{\dagger}
-v^2W_{\mu}W^{\mu} \nonumber \\
&&-\frac{1}{2e^2(\Sigma)} G_{MN}G^{MN}\Bigr]\,,
\end{eqnarray}
with
\beq
 W_{\mu}={i \over 2v^2}\left[\hat{D}_\mu H_{L}H_{L}^\dagger
-H_{L}(\hat{D}_\mu H_{L})^\dagger+(L\leftrightarrow R)\right].
\eeq
Eliminating $W_\mu$, we obtain 
the following expression for the integrand of the 
effective Lagrangian after some simplification 
\begin{equation}\label{eq:lagr1}
\oper{L}_{\rm eff} = \frac{1}{2v^2}\int_{-\infty}^{\infty} 
dy\, \Tr\Bigl[\oper{D}_{\mu}H_{ab}\oper{D}^{\mu}H_{ba}
\Bigr]\,,
\end{equation}
where we defined fields $H_{ab}$ with the label $ab$ of adjoint 
representation of the flavor gauge group $SU(N)_{L+R+c}$ and 
the covariant derivative as
\begin{eqnarray}
 \D_\mu H_{ab}=\partial_{\mu}H_{ab}+i[A_\mu, H_{ab}], 
\quad H_{ab}\equiv H_a^\dagger H_b, \quad a,b=L,R.
\end{eqnarray}
The full description of the procedure to obtain the effective Lagrangian in the closed form can be found in \cite{arxiv}.
%In  \ref{app2}, 
%we will describe fully the procedure to derive the effective 
%Lagrangian by substituting (\ref{eq:solU1_SY3}) and 
%(\ref{eq:solU2_SY3}) and rewriting in terms of moduli 
%fields $\hat{x}$ and $U$. 
Here we merely state the result:
\begin{align}
 \oper{L}_{1,{\rm eff}} & = \frac{v^2}{2m}
\Tr\biggl[\oper{D}_{\mu}\hat x\,
 \frac{\cosh(\oper{L}_{\hat x})-1}{\oper{L}_{\hat x}^2
\sinh(\oper{L}_{\hat x})}
\ln\biggl(\frac{1+\tanh(\oper{L}_{\hat x})}
{1-\tanh(\oper{L}_{\hat x})}\biggr)(\oper{D}^{\mu}\hat x) 
\nonumber \\
& +U^{\dagger}\oper{D}_{\mu}U\,
\frac{\cosh(\oper{L}_{\hat x})-1}{\oper{L}_{\hat x}
\sinh(\oper{L}_{\hat x})}
\ln\biggl(\frac{1+\tanh(\oper{L}_{\hat x})}
{1-\tanh(\oper{L}_{\hat x})}\biggr)(\oper{D}^{\mu}\hat x) 
\nonumber \\
&+\frac{1}{2}\oper{D}_{\mu}U^{\dagger}U\frac{1}
{\tanh(\oper{L}_{\hat x})}\ln\biggl(
\frac{1+\tanh(\oper{L}_{\hat x})}{1-\tanh(\oper{L}_{\hat x})}
\biggr)
(U^{\dagger}\oper{D}^{\mu}U)\biggr],
\label{eq:result}  
\end{align}
where 
\begin{equation}
{\cal L}_A(B)=[A,B]
\label{eq:lie_derivative}
\end{equation} 
is a Lie derivative with respect to $A$.
The covariant derivative $\oper{D}_\mu $  is defined by 
\beq
\oper{D}_\mu U = \p_\mu U + i \left[A_{\mu}, U\right].
\eeq

%The above result suggests that the chiral fields $U(x^\mu)$ 
%and hermitian fields $\hat x(x^{\mu})$ are 
%in the adjoint representation of $SU(N)_{L+R}$. 
%Let us now examine the transformation property of 
%$U$ and $\hat x$ under the $SU(N)_{L+R}$ flavor gauge 
%transformation on the domain wall background in order to 
%demonstrate that they are in the adjoint representation. 
%The domain wall solution only preserves the diagonal subgroup 
%$SU(N)_{L+R+c}$. Eqs.(\ref{eq:transf_H}) and 
%(\ref{eq:transf_sigma}) shows 
%the fields transform under the $SU(N)_{L+R+c}$ 
%transformations ${\cal U}$ as 
%\begin{equation}
%H'_L={\cal U} H_L {\cal U}^\dagger, \quad 
%H'_R={\cal U} H_R {\cal U}^\dagger, \quad 
%\Sigma'={\cal U} \Sigma {\cal U}^\dagger. 
%\end{equation} 
%Eqs.(\ref{eq:solU1_SY}) and (\ref{eq:solU2_SY}) shows that 
%\begin{equation}
%S'={\cal U} S {\cal U}^\dagger, 
%\quad e^{\phi'}={\cal U}e^\phi {\cal U}^\dagger, \quad 
%\Omega'={\cal U}\Omega {\cal U}^\dagger.
%\end{equation}
%The complex moduli $e^\phi$ is decomposed into hermitian part 
%$e^{\hat x}$ and unitary part $U$ in Eq.(\ref{eq:xu_decomp}). 
%Since we can express 
%$e^{2\hat x}=e^{\phi} e^{{\phi}^\dagger}$, 
%and $U=e^{-\phi}e^{\hat x}$, 
%we find that they transform as adjoint representations 
%\begin{equation}
%%e^{2\hat x'}
%%=e^{\phi'} e^{{\phi'}^\dagger}
%%={\cal U} e^{\phi} e^{{\phi^\dagger}}{\cal U}^\dagger
%={\cal U}e^{2\hat x} {\cal U}^\dagger, 
%\quad U'={\cal U} U {\cal U}^\dagger. 
%\end{equation}

By expanding (\ref{eq:result}), we here illustrate nonlinear 
interactions of $\hat{x}$ up to fourth orders in the 
fluctuations $\hat x$ and $U-{\bf 1}$ 
\begin{eqnarray}
\Lag_{1,\rm eff}&=&\frac{v^2}{2m}
\Tr\Bigr(\oper{D}_{\mu}U^{\dagger}\oper{D}^{\mu}U
+\oper{D}_{\mu}\hat x\oper{D}^{\mu}\hat x 
+U^{\dagger}\oper{D}_{\mu}U\comm{\hat x}{\oper{D}^{\mu}\hat x}
\nonumber \\
&&
-{1 \over 12}\comm{\oper{D}_{\mu}\hat x}{\hat x}
\comm{\hat x}{\oper{D}^{\mu}\hat x}
+\frac{1}{3}[\oper{D}_{\mu}U^\dagger U, \hat x]
[\hat x, U^\dagger \oper{D}^{\mu}U]
+\cdots \Bigr).
\label{eq:4thorder_nonlinear}
\end{eqnarray}

\begin{figure}[ht]
\begin{center}
\includegraphics[width=8cm]{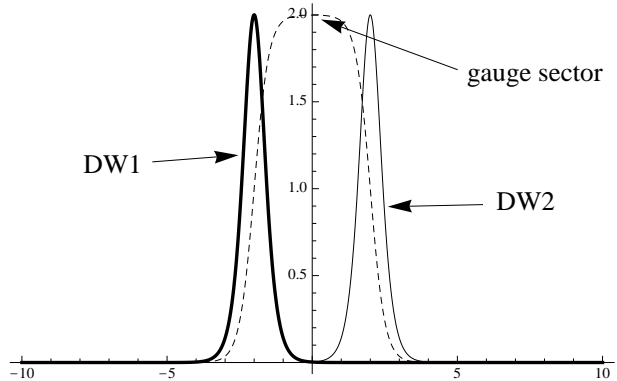}
\caption{The wave functions of the zero modes. 
DW1 and DW2 stand for the wave functions 
of the massless matter fields of the $i=1$ 
domain wall and $i=2$ domain wall, respectively 
for strong gauge coupling limit $g_i=\infty$ and 
$m_i=1$. 
The gauge fields are localized between the domain walls.}
\label{fig:schematic}
\end{center}
\end{figure}

In a similar way to Eq.(\ref{eq:integral4Dcoupling}), we 
can define the (3+1)-dimensional non-Abelian gauge 
coupling $e_{4}$ 
by integrating (\ref{Lag_gChi2}) and find 
\beq
\frac{1}{2e_{4}^2} 
=\int dy \frac{1}{2e^2(\Sigma)} 
%= \frac{\lambda}{2}\int dy \left(\frac{\Tr\Sigma_1}{Nm_1} 
%- \frac{\Sigma_2}{m_2}\right)
= \lambda(y_2-y_1)\,,
\label{eq:4d_NA_gauge_coupling}
\eeq
where $y_i$ is the wall position for the $i$-th domain wall 
sector. 
Summarizing, we obtain the following effective Lagrangian
\beq
\Lag_{\rm eff} = 
\Lag_{1, {\rm eff}} + 
\Lag_{2, {\rm eff}}- \frac{1}{2e_{4}^2}
\Tr \Bigl[G_{\mu\nu}G^{\mu\nu}\Bigr],
\label{eq:full_eff_Lag}
\eeq
where $\Lag_{2, {\rm eff}}$ is given in (\ref{eq:effLagAbelian}).
This is the main result.
%Let us note that we have chosen the coincident walls.
We have succeeded in constructing the low energy 
effective theory in which the matter fields 
(the chiral fields) and the non-Abelian gauge 
fields are localized with the non-trivial interaction.
We show the profile of "wave functions" of localized 
massless gauge field and massless matter fields 
as functions of the coordinate $y$ of the extra 
dimension in Fig.~\ref{fig:schematic}.

As is seen from Eq.(\ref{eq:4thorder_nonlinear}), 
the flavor gauge symmetry $SU(N)_{L+R+c}$ is further 
(partly) broken and the corresponding gauge field 
$A_\mu$ becomes massive, when the fluctuation 
$\phi = e^{\hat x} U$ develops non-zero vacuum expectation values.
Especially, $\hat x$ is interesting because 
its non-vanishing (diagonal) values of the fluctuation 
has the physical meaning as the separation 
between walls away from the coincident case. 
For instance, if all the walls are separated, 
$SU(N)_{L+R+c}$ is spontaneously broken to
the maximal $U(1)$ subgroup $U(1)^{N-1}$. 
However, if $r$ walls are still coincident 
and all other walls are separated, we have an unbroken 
gauge symmetry $SU(r)\times U(1)^{N-r+1}$.
Then, a part of the pseudo-NG modes $\hat x$ turn to NG modes associated with
the further symmetry breaking $SU(N)_{L+R+c} \to SU(r) \times U(1)^{N-r+1}$, so that
the total number of zero modes\footnote{In Ref.\cite{ShYu2} the authors argued that the non-Abelian clouds
spreading between walls become massive contrary to the results of \cite{EtFuNiOhSa}.} is preserved \cite{EtFuNiOhSa}.
These new NG modes, called the non-Abelian cloud, spread between the separated domain walls \cite{EtFuNiOhSa}.
The flavor gauge fields eat the non-Abelian cloud and get masses which are proportional
to the separation of the domain walls. This is the Higgs mechanism in our model.
This geometrical understanding of the Higgs mechanism is quite similar to D-brane systems
in superstring theory. So our domain wall system provides a genuine prototype
of field theoretical D3-branes.

\bigskip
\section{Conclusions and discussion}\label{sc:6}
In this work we have successfully localized both massless 
non-Abelian gauge fields and massless matter fields 
in non-trivial representation of the gauge group. 
%We first considered a (4+1)-dimensional $U(N)$ gauge theory 
%with additional $SU(N)_L\times SU(N)_R\times U(1)_{A}$~flavor 
%symmetry. 
%We introduced the flavor gauge field for the diagonal 
%flavor group $SU(N)_{L+R}$, which is unbroken in the 
%coincident wall background. 
%The flavor gauge fields are localized on the wall by 
%introducing the scalar-field-dependent gauge coupling 
%function. 
Then we studied the low-energy effective Lagrangian 
and showed that massless localized matter fields 
interact minimally with localized 
$SU(N)_{L+R}$ gauge field as adjoint representations. 
Moreover, full nonlinear interaction between the moduli 
containing up to the second derivatives, was worked out.
%The field-dependent gauge coupling function is naturally 
%realized in supersymmetric gauge theories 
%using the so-called prepotential. 
%For this reason, we also explored bosonic part of 
%${\cal N}=1$ supersymmetric extension of our model.

Main result of this paper is the effective Lagrangian 
(\ref{eq:full_eff_Lag}). The moduli field $U$ appearing 
in the effective theory, is a chiral $N\times N$ matrix 
field like a pion, since it is a NG boson of spontaneously 
broken chiral symmetry. 
Other moduli in (\ref{eq:full_eff_Lag}), denoted by 
$N\times N$ Hermitian matrix $\hat x$, has the physical 
meaning of positions of $N$ domain walls as its diagonal 
elements. 
We argued that the fluctuations of moduli field $\hat x$, 
can develop VEV corresponding to splitting of walls, 
and the Higgs mechanism will occur as a result.
Namely, the flavor gauge fields get masses by eating 
the non-Abelian cloud. 
Therefore, in this model, Higgs mechanism has a 
geometrical origin like low energy effective theories on D-branes
in superstring theory. 

Amongst the possible future investigations, we would like 
to study non-coincident solution to further clarify this 
geometrical Higgs mechanism. %For further discussions see \cite{arxiv}. 

\end{document}